\documentclass[journal = jpccck,manuscript=article]{achemso}
\usepackage{graphicx}
\usepackage{xcolor}
\usepackage{nicefrac}
\usepackage[utf8]{inputenc}
\usepackage{array}

\title{Monolayer and bilayer perfluoropentacene on Cu(111)}
\author{S.  Smalley}
\affiliation{Jeremiah Horrocks Institute for Mathematics, Physics and Astronomy, University of Central Lancashire, Preston PR1 2HE UK}
\author{P.  Darancet}
\affiliation{Center for Nanoscale Materials, Argonne National Laboratory, Il 60439, USA}
\author{J.  R.  Guest}
\affiliation{Center for Nanoscale Materials, Argonne National Laboratory, Il 60439, USA}
\author{J.  A.  Smerdon}
\email{jsmerdon@uclan.ac.uk}
\affiliation{Jeremiah Horrocks Institute for Mathematics, Physics and Astronomy, University of Central Lancashire, Preston PR1 2HE UK}
\begin{document}
\begin{abstract}
Perfluoropentacene (PFP), an $n$-type organic semiconductor, is deposited at monolayer and bilayer coverage on Cu(111).  Scanning tunneling microscopy at various bias voltages is used to investigate the geometric and electronic structure of the layer.  The appearances of the first layer and second layer differ, probably because of perturbation to the first layer electronic structure by the substrate.  This has been previously observed for pentacene (Pn), the isostructural $p$-type organic semiconductor.  The PFP film has a unit cell of (4, -3,3 4) relative to the substrate, which is larger than that of Pn/Cu(111), representing a half-integer increment in each direction.
\end{abstract}

\maketitle

\section{INTRODUCTION}
Perfluoropentacene (PFP) is a linearly-bonded chain of five benzene rings, terminated with F atoms \cite{sakamoto2004perfluoropentacene}.  Its cousin pentacene (Pn), which is terminated with H atoms, is a $p$-type organic semiconductor \cite{sakamoto2004perfluoropentacene,singh2005high,daraktchiev2005ultrathin} that is used in organic solar cells and other molecular electronics due to its high hole mobility \cite{klauk2002high}, stability during atmospheric operation \cite{yan2009ambipolar} and the strength of molecule--molecule interactions which make it suitable for self assembly applications and high quality crystal growth.

The electronic structure of organic semiconductors is strongly affected by the ordering of the molecular crystal \cite{kolata2014molecular}.  The electronic behavior of organic \textit{devices} is to a large extent governed by, or at least subject to, the quality of the interfaces between organic material and metallic interconnects.  Molecular adsorption on ideal metal surfaces provides useful model systems for the study of both of these types of phenomena.  For example, Pn, a simple oblong-shaped molecule, has complicated behavior and several different growth modes arising from the competing interactions between neighboring molecules and the substrate on which they are adsorbed \cite{kawai2011systematic,rinn2017interfacial,kang2003pi}.  Organic semiconductors have several benefits compared to inorganic semiconductors, such as flexibility, transparency and non-toxicity \cite{schwarze2016band}.  Exploiting their unique properties such as low film thickness and accompanying short transport distances can yield devices with short response times for high frequency applications \cite{forrest2004path}.  Tuning of the HOMO and LUMO energy levels in organic systems, akin to engineering the valence and conduction bands in traditional semiconducting systems, is key to optimising device efficiency.  Halogenation and fluorination in particular have been explored as mechanisms to achieve this \cite{anger2012photoluminescence,shen2018bridging,kim2015difluorinated}.  While development of organic electronic and optoelectronics has brought some devices to market, performance is still a barrier to many applications \cite{otero2017electronic,schwarze2016band}.  The performance of a molecular device is dependent on a range of interacting properties such as substrate surface contact resistance, adsorption geometry \cite{toyoda2011density} and the degree of order.  

Coinage metals Au, Ag and Cu \cite{lu2016pentacene,gotzen2010structural,eremtchenko2005formation} are of particular interest as they are widely used in electronics \cite{smerdon2011monolayer,soe2009direct}.  Of these, the (111) surface of Cu has the strongest interaction with Pn, followed by Ag(111) and then Au(111) \cite{lu2016pentacene}.  In all cases, though to different degrees, the diffusion of sub-monolayer Pn is observed \cite{lo2013comparative}.  Atop Cu(111), Pn interacts strongly, with substantial charge transfer leading to large differences in appearance and electronic structure upon adsorption \cite{smerdon2011monolayer}.  On weakly interacting surfaces, intermolecular interactions usually dominate, resulting in Pn adopting a standing herringbone structure \cite{kawai2011systematic}.  On Cu(111), random tiling, herringbone close-packing and further close-packed linear islands stabilized by bilayer (BL) growth have all been observed at different coverages and deposition temperatures \cite{smerdon2011monolayer,kawai2011systematic}.  In an earlier study, we demonstrated the electronic behavior of Pn as the semiconductor half of a molecular Schottky diode, with C$_{60}$ doped via its interaction with Cu(111) as the metallic contact \cite{smerdon2016large}.

Studies of molecular adsorption have focused on the first two layers for the reason that these layers contain molecule-surface and molecule-molecule interactions.  The properties of subsequent layers are not expected to vary considerably from those of the second layer, since each molecule interacts directly only with its nearest neighbors. Indirect influence from the substrate can be observed in later layers only if transmitted through the earlier layers. A good example of this is when the first and second layer are stabilized via interaction with the substrate in a structure that is not stable in the bulk molecular crystal, for example, $\pi$-stacked flat-lying Pn \cite{kang2003pi,smerdon2011monolayer}.  In this example, at some thickness greater than 1 ML, the film transforms to a bulk-like phase, though the critical thickness for this transition seems to be governed by complex factors. It is not known if the first layer is included in the transformation.

As a result of fluorination, PFP differs in one additional major aspect from Pn: it is an $n$-type semiconductor \cite{sakamoto2004perfluoropentacene}.  Both Pn and PFP adopt herringbone packing in the bulk, albeit with different lattice parameters and angles \cite{sakamoto2004perfluoropentacene,smerdon2011monolayer}.  In principle, the structural similarity between bulk Pn and bulk PFP suggests the possibility of intermixed molecular crystals, and therefore effective doping of one by the other.  Some studies show apparently effective intermixing, though only for 1:1 mixtures \cite{hinderhofer2011structure,rinn2017interfacial}.  A recent study of epitaxial growth of PFP on pentacene single crystals showing a sharp interface indicates that there is no spontaneous intermixing at room temperature \cite{Nakayama2019}. PFP has also been studied on Au(111), Ag(111) and Cu(111) metals at monolayer (ML), bilayer (BL) and multilayer coverages and in general adsorbs more weakly than Pn \cite{lo2013comparative,goiri2014self}.  It has also been investigated as a ML and BL atop SiO$_{2}$ and graphene on SiO$_{2}$ \cite{salzmann2012epitaxial}, forming a vertical herringbone and a planar structure, respectively.  These studies highlight the importance of in-depth characterization of the the interface structures due to their influence on molecule-molecule interactions and the resultant electronic properties of the system.

The structure of PFP/Cu(111) has been investigated at room temperature (RT) and after annealing at both BL and ML coverage \cite{glowatzki2012impact,koch2008adsorption}.  Glowatzki \textit{et al.} report that up to and including ML coverage, the PFP adsorbs in a disordered fashion at RT and behaves as a 2D liquid, unresolved via STM due to its rapid diffusion across the surface (other than at step edges and defects \cite{glowatzki2012impact}).  The 2D liquid behavior reported at RT is consistent with the behavior of Pn on Ag(111) at RT \cite{dougherty2008variable,glowatzki2012impact}.

Variable temperature studies of PFP on Ag(111) have characterized it as a structurally strained system due to the substrate geometry.  At RT, PFP forms an ordered layer of end-to-end rows of molecules that shows a moir\'e pattern which repeats every 8 molecular rows.  The moir\'e pattern shows that in this case each molecule has a slightly different, strained, registry with the substrate.  When cooled to 90 K, PFP/Ag(111) reorders to a layer of slightly lower density, in which there is no moir\'e pattern but instead a structure in which the row structure is broken by a dislocation every 6 molecules \cite{goiri2012understanding}.  This dislocation pattern lifts the strain and places each molecule in the same registry with the substrate.  PFP forms 6 domains on this system at each temperature, with 3$\times$ rotational degeneracy due to the substrate symmetry and 2$\times$ chiral degeneracy due to the moir\'e pattern at RT, or alternatively the row kinking at 90 K.

In general, two competing interactions dictate PFP adsorption: the strong attraction of the central acene rings to the substrate and a repulsion between F atoms \cite{franco2018resolving} and the substrate.  This phenomenon introduces a bowl-like adsorption conformation with upwardly-curved molecule edges \cite{franco2018resolving,toyoda2011density,koch2008adsorption} and suggests that the molecule--substrate adsorption distance is increased by the shape of the 2p orbital of terminal F atoms compared to the 1s orbital of terminal H in pentacene \cite{toyoda2011density}.  An increased molecule--substrate separation leads to reduced orbital hybridisation and reduced HOMO/LUMO broadening as a result \cite{toyoda2011density}.

DFT simulations have explored variably fluorinated Pn molecules, in particular F$_4$Pen, using F substitution as a tool to tailor the electron injection barrier; the addition of F lowers the molecular HOMO energy \cite{inoue2005organic}.  Partial fluorination has been shown to improve crystalline ordering and transport efficiencies in pentacene derivatives \cite{shen2018bridging,kim2015difluorinated}.  PFP has been used in OTFTs with carrier mobilities comparable with Pn OTFTs \cite{inoue2005organic} but, as of writing, $p$-type OTFTs are outperforming $n$-type OTFTs in this area \cite{paterson2018recent}.  Development of equally efficient $n$-type OTFTs is necessitated by the use of both in CMOS applications \cite{kumar2014organic}, bipolar transistors \cite{lo2013comparative} and complementary or ambipolar circuits \cite{inoue2005organic} which are of particular interest in efficient logic gate design \cite{ben2009novel} and opto-electronic light-emitting devices \cite{rost2004-apl}.  Ambipolar devices are those capable of both electron and hole transport, and their operation can be facilitated by the presence of both electron donor and acceptor moieties within a molecule or crystal structure  \cite{duan2011strategies}.

In this work we investigate the ML and BL structures of PFP on Cu(111) at 50 K.  Elucidating the geometry of these systems is an important step towards understanding the interactions that govern their electronic properties and therefore applications in devices.

\section{METHODS}
All measurements are taken in a commercial Omicron variable temperature STM under ultra-high vacuum (UHV) conditions with base pressure $5.0\times10^{-10}$ mbar in the analysis chamber and $6.0\times10^{-9}$ mbar in the preparation chamber.  The Cu(111) crystal was prepared through overnight simultaneous sputter/anneal preparations with 1.5 keV Ar$^+$ ions at 900 K followed by a shorter 30 minute anneal at 900 K prior to scanning.

PFP was evaporated from a Dodecon four-cell organic molecular-beam epitaxy source at 480 K.  Prior to each deposition the source was out-gassed at 450 K.  The Cu(111) sample was at RT throughout the 10 minute deposition process.  This resulted in sub-ML coverage and repetition produced a partial BL.  Both of these systems were characterised using STM with a mechanically cut PtIr (90:10) tip at 50 K.  We use the convention of sample bias when referring to bias voltage.

Scanning tunneling microscopy is performed in `constant current' mode.  The signal in the $I$ channel is therefore related to the error signal for the $Z$ channel, and sometimes provides images with clearer detail.  Where these are presented, they are identified as $I_{TC}(x,y)$ images.

\section{RESULTS AND DISCUSSION}

Perfluoropentacene was deposited atop clean Cu(111) to bilayer coverage, as described above, before insertion into the STM and cooling to 50 K.  Figure \ref{dataimages}(a) shows simultaneous resolution of both BL and ML.  In Figure \ref{dataimages}(c) two BL islands are observed, one on each side of a Cu step edge.  

Monolayer and BL are structurally similar; both have the same 2D lattice parameters.  In the $I_{TC}(x,y)$ image in Figure \ref{dataimages}(b), lines are superimposed on molecules in each layer, with the blue series over the long axes of ML molecules and the green series over similar locations in BL molecules.  The half-integer relationship between these relative position of these series demonstrates the existence of a lateral offset between layers, as also observed for Pn/Cu(111) \cite{smerdon2011monolayer}.

\subsection{Monolayer and bilayer domains}
\begin{figure}
 \includegraphics[width=1\textwidth]{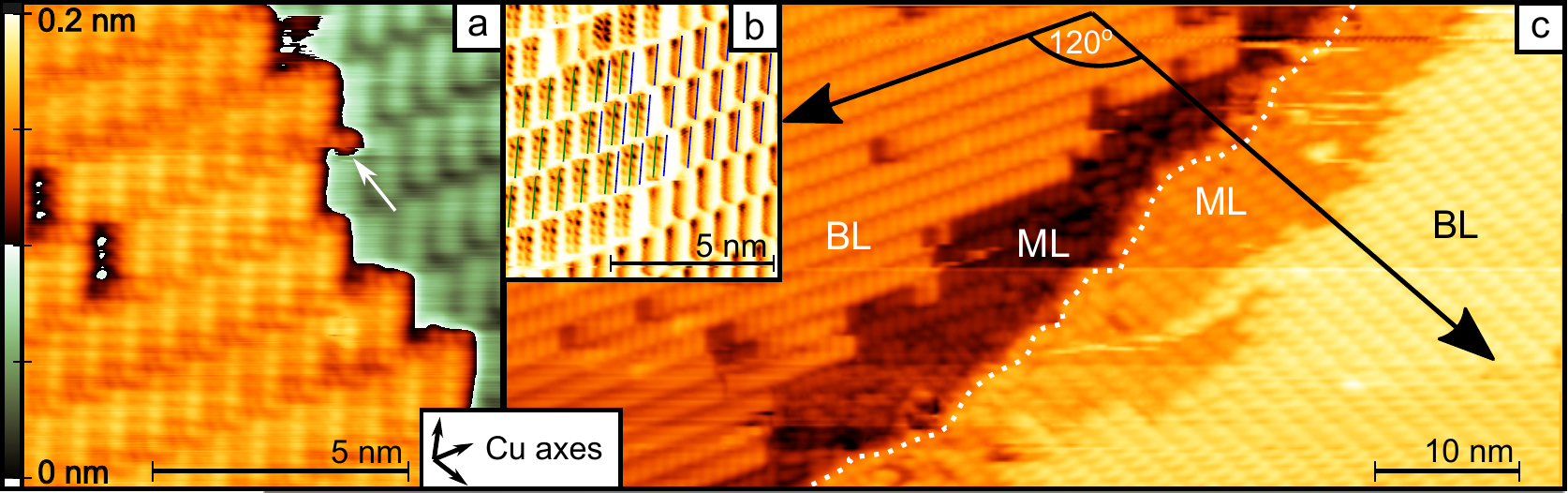}
 \caption{a) STM topograph of a PFP bilayer boundary region ($I_T=100\textrm{pA}, V_B=-2.5$V).  The bifurcated lobes of the HOMO of bilayer molecules are evident on the left hand upper \nicefrac{1}{3} of the image.  A tip change occurs \nicefrac{1}{3} of the way down the image, with the result that the molecules below the change have a pronounced bright lobed feature, as in the results of Glowatzki \textit{et al.} \cite{glowatzki2012impact}.  An arrow indicates a molecule that has apparently disappeared following the change.  Further discussion in text.  b) A $I_{TC}(x,y)$ image over a similar BL/ML step.  The difference between the orbital shapes is apparent.  The long axes of the molecules in the BL and ML are marked to illustrate the half-integer unit cell offset between them ($I_T=100\textrm{pA}, V_B=-2.5$V).  c) Two domains of the PFP BL separated by a Cu terrace ($I_T=100\textrm{pA}, V_B=-3$V).}
 \label{dataimages}
\end{figure}

The behavior of Pn atop Cu(111) and the similar anisotropic nature of PFP lead us to expect that the PFP molecules adsorb with a long axis coincident with the substrate NN (nearest-neighbor) axis.  We confirm this independently by comparison between images such as in Figure \ref{dataimages} and images of C$_{60}$ adsorbed atop the same substrate collected contemporaneously (see Supporting Information (SI)).

The threefold nature of the substrate means that molecules within a domain are aligned in one of three directions, angled 120$^{\circ}$ apart, as seen in the angular separation of two domains marked in Figure \ref{dataimages}(c).  While, similarly to the behavior of Pn adsorbed on Cu(111), the PFP molecular long axes are aligned with substrate axes, in contrast to the case of Pn, the \emph{domains} are not; lines drawn connecting equivalent sites in molecules in a domain are \emph{not} parallel with substrate axes.  Molecules within a domain therefore have staggered junctions with their neighbours, such that equivalent sites in adjacent molecules occupy sites on \emph{adjacent} NN rows of substrate atoms.  From the combination of 2-fold mirror symmetry and 3-fold rotational symmetry, we expect a total of 6 domain types at both ML and BL coverage, similarly to the case for adsorption atop Ag(111) \cite{goiri2012understanding}.

Comparing the appearance of the molecules in Figure \ref{dataimages}(a) with the orbital shapes presented in Figure \ref{structureimages}(a), BL molecules apparently tilt about their long axis, though -- depending on scan parameters -- this tilt is not always resolved.  

Glowatzki \textit{et al.} report diffusion of PFP across the Cu(111) surface at RT for coverage less than a bilayer.  Clearly, an ordered bilayer with domain symmetry originating in the substrate requires an ordered monolayer.  They suggest that the second layer locks the first layer into registry, as has been previously reported for Pn/Ag(111) at RT \cite{eremtchenko2005formation} and Pn/Cu(111) \cite{smerdon2011monolayer}.  At 50 K, we directly resolve the ML alongside the BL as in Figures \ref{dataimages}(a) and \ref{dataimages}(b).  Resolution of the ML is also possible independent of the BL (as presented in SI).  In Figure \ref{dataimages}(b), the structure of both the ML and BL break down near the Cu step edge.  Disordered regions are visible in the ML regions, but are not observed in the BL.

\subsection{Electronic structure}

Scanning is performed at negative bias, and thus occupied states of the PFP are imaged.  In Figures \ref{dataimages}(a) and (b), the well-defined lobes relating to the HOMO structure of PFP are evident in the BL region of the topography, while 
 the ML is seen to consist of featureless 1D rods independent of tunneling parameters; the different appearance of the BL and ML molecules indicates a difference in electronic structure.  Simulated HOMO and LUMO charge isosurfaces are presented alongside the bias dependent STM images of the BL molecules in Figure \ref{structureimages}(a).

Following an earlier analysis of Pn/Cu(111) \cite{smerdon2011monolayer}, we suggest that the first layer of PFP molecules participates in charge transfer with the substrate.  This behavior arises when the LUMO of the molecular overlayer is broadened through adsorption, including the Fermi energy of the substrate \cite{heimel2013charged,otero2017electronic}.  This results in partial filling of the LUMO and charge transfer between the substrate and the ML, leading to a Fermi surface in the molecular layer and a degree of metallic conduction \cite{tzeng2007covalent,pai2010optimal,wang2004rotation}.

\subsection{Adsorption site lattice determination}

For some molecular adsorption systems, the adsorption site lattice (ASL) is difficult to establish. Confounding factors include the presence of an observed moir\'e pattern (as in the case for RT PFP adsorption atop Ag(111) \cite{goiri2012understanding}); a moir\'e pattern indicates that each molecule within the moir\'e superstructure unit cell has a different registry with the substrate, leading to a slightly different adsorption conformation and/or electronic structure.

In addition, the usually vastly different tunneling parameters required for substrate atomic resolution and molecular resolution mean it is impossible to capture both in a single image. This means it is usually not possible to directly determine the precise adsorption site. However, lack of knowledge of the precise adsorption site does not necessarily imply lack of knowledge of the precise ASL.  

In the present case, the precise PFP adsorption site on the surface cannot be directly determined through our STM data for the reason above: we do not have atomic resolution on the clean Cu(111) surface.  We refer to a study of Pn atop Cu(111) for guidance \cite{lagoute2004manipulation}.  In the cited article, Cu adatoms were deposited simultaneously with single Pn molecules.  The Cu adatoms adopted sites consistent with a Cu epitaxial adlayer, so, in reference to their locations, the substrate lattice could be determined.  This led to an atomically precise determination of the Pn adsorption site.  Our assumption is that PFP adopts the same adsorption site, which is with the molecule center over an hcp hollow site.  This adsorption pattern is also consistent with similar polyaromatic molecules on the close packed surfaces of coinage metals. It is also consistent with the $\pi$ system of the PFP ML being the dominant component in the molecule--substrate interaction, as was observed with Pn.  As the $\pi$ bond arrangement is the same for both PFP and Pn, they likely adsorb similarly atop Cu(111).

As mentioned above, by comparison with contemporaneous data of C$_{60}$ atop the same Cu crystal, we are able to directly determine the orientation of PFP molecules: as for Pn, they adsorb with the molecular long axis parallel to substrate NN axes.

Koch \textit{et al.} find that PFP distorts upon adsorption of Cu(111) so that the C backbone of the molecule is closer to the substrate than the terminating F atoms \cite{koch2008adsorption}.

The ASL is determined from the STM data.  We find it by simply comparing the dimensions from STM data to those available given the substrate constraints on geometry. Specifically, the lack of a moir\'e pattern indicates that the ASL unit cell has integer vectors with substrate unit vectors.  Our previous work on Pn/Cu(111)\cite{smerdon2011monolayer} shows this idea in deeper detail by comparing possible competing lattices.  It is clear from this earlier work that competing lattices would produce vastly different STM data.  It turns out that, for relatively small molecules such as PFP or Pn and in the absence of confounding factors, there is only one ASL that is permitted by the substrate geometry.  For Pn/Cu(111), this is (3 3,0 6).  For PFP/Cu(111), it is (4 -3,3 4).

\begin{figure}
 \includegraphics[width=1\textwidth]{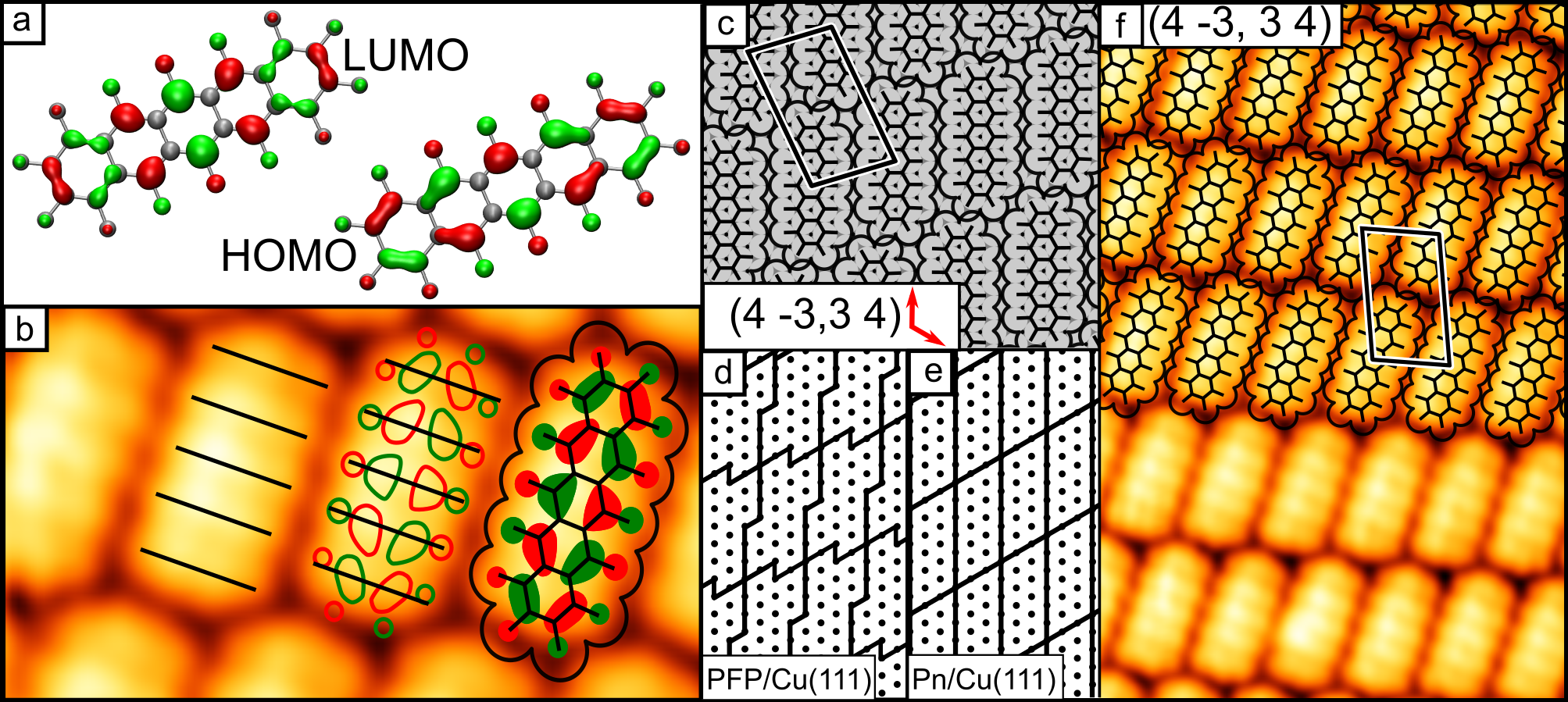}
 \caption{a) Charge isosurfaces of PFP HOMO/LUMO states;  \cite{chen2005effect} b) A structural model of the PFP molecule scaled to the HOMO identified in the the data ($I_T=50 \textrm{pA}, V_B=-2 $V);  
 c) The (4 -3,3 4) ASL for the PFP ML and BL as determined from the data presented as a wire model with the van der Waals radii of each molecule.  The unit cell as reported by Glowatzki is shown in red for comparison;  d) The `kinked' (4 -3,3 4) ASL of PFP/Cu(111) vs e) the `straight' ASL of Pn/Cu(111) \cite{smerdon2011monolayer}; f) The PFP model is superimposed atop high resolution data of the BL ($I_T=50 \textrm{pA}, V_B=-2 $V).}
 \label{structureimages}
\end{figure}

High resolution images with superimposed elements to aid the reader are shown in Figures \ref{structureimages}(b) and \ref{structureimages}(f). Under these tunneling conditions, the molecules appear as symmetrical features with 5 prominent lobes along their length.  The first lobe in one molecule lines up with the second in the next, and so on.  This provides another constraint in determining the relationship between one molecule and its neighbors within a row.

The side-by-side alignment of submolecular features in adjacent molecules indicates inter-molecular side-by-side alignment of terminal F atoms, which contrasts with the observation of interleaved F atoms reported for PFP/Ag(111) \cite{goiri2012understanding}.  The (4 -3,3 4) ASL mentioned above, and presented in Figure \ref{structureimages}(c) and superimposed onto the data in Figure \ref{structureimages}(f) produces this F arrangement.

The (3 3,0 6) Pn/Cu(111) ASL is a denser structure than the PFP/Cu(111) ASL, with the unit cell smaller by \nicefrac{1}{2} of a NN in each direction.  Each molecule is separated from its neighbors by several Cu NN distances and 
the edges of the Wigner-Seitz unit cell coincide with primary Cu(111) surface axes. The structures are compared in Figure \ref{structureimages}(d,e).

The larger van der Waals radii of the F atoms in PFP force a looser (4 -3,3 4) packing.  The half-integer nature of the expansion forces the F atoms into approximate side-by-side alignment \textit{and also} means that every adjacent molecule is on a different set of substrate rows.  This means that there are no straight lines of molecules.  Every line is kinked compared to the substrate, as shown in Figure \ref{structureimages}(c).  
%
\begin{table}[]
  \centering
  \begin{tabular}{c m{1.1cm} m{1.1cm} m{1.1cm} m{1.5cm} m{1.5cm}}
     system & \textit{a} (nm) & \textit{b} (nm) & $\theta$ ($^\circ$) & Area (nm$^2$) & Density (nm$^{-2}$) \\ \hline
Glowatzki PFP description   & 1.0 & 1.75 & 92  & 1.75     & 0.57        
        \\
Our PFP description   & 1.04 & 1.75 & 99.2  & 1.81     & 0.55        \\
Pn    & 0.77 & 1.54 & 60    & 1.02	    & 0.98  \\   
  \end{tabular}
  
  \caption{The unit cells of the PFP/Cu(111) as reported here, in the work of Glowatzki \textit{et al.} \cite{glowatzki2012impact} and the Pn/Cu(111) system reported by Smerdon \textit{et al.} \cite{smerdon2011monolayer}.}
  \label{tableoflattices}
\end{table}

The (4 -3,3 4) ASL lattice parameters are therefore $a$ = 1.04 nm, $b$ = 1.76 nm, $\theta$ = 99.2$^\circ$, as presented alongside the Pn/Cu(111) in the table above.  We do not present associated errors on these values because they are derived from the Cu lattice constant.  These measurements show good agreement to those observed by 
Glowatzki \textit{et al.} \cite{glowatzki2012impact}, who use LEED and STM to measure the structure, reporting the surface mesh of the bilayer as $a$ = (1.75 $\pm$ 0.05) nm, $b$ = (1.0 $\pm$ 0.1) nm, $\theta$ = (92 $\pm$ 1)$^\circ$.  
The major difference between our description and this is the angle of the unit cell, likely due in part to the thermal drift that usually confounds STM measurements.  The Glowatzki model does not take the approach of making arguments based on the structure of the substrate, instead presenting direct measurements from STM and LEED.

Glowatzki \emph{et al.} propose that the BL molecules have a tilt about the molecular long axis and suggest that this originates in an offset between the BL and the ML, similar to the bilayer structure observed on Ag(111) \cite{gotzen2010structural}, our earlier study of Pn/Cu(111) \cite{smerdon2011monolayer} and the data under discussion, as evidenced by the half-unit cell offset shown in Figure 1(b).  This structure would serve as an intermediary between the flat monolayer and the herringbone stacking seen in the bulk and in thick multilayers and facilitates maximum contact between the electronegative F atoms of one layer and the $\pi$ system of the other.

\section{CONCLUSIONS}
The PFP monolayer and bilayer structures atop Cu(111) have been elucidated.  Monolayer and bilayer PFP adopts a flat lying (4 -3,3 4) ASL at 50 K. Based on the difference between simultaneous observations of molecules in the first and second layers, it is clear that the electronic structure is modified, which we propose is due to charge transfer from the metal substrate in a manner similar to Pn on Cu(111) \cite{smerdon2011monolayer}.  The BL exhibits clearly resolved orbitals at a range of scanning biases that mimic the HOMO and LUMO features expected of an isolated molecule.  This suggests the semiconducting properties of the BL molecules are unchanged.

The row structure is not aligned along substrate NN rows, as for Pn on Cu(111) \cite{smerdon2011monolayer,lo2013comparative}.  The larger `footprint' of PFP forces the unit cell to increase in size by \nicefrac{1}{2} a Cu NN distance in each direction.  To maintain the same registry between molecules and surface sites, each adjacent row of molecules is displaced by a substrate atomic row in both directions, leading to an overall kinked appearance of the molecular rows.

Bulk PFP has a larger unit cell than bulk Pn and this difference is exaggerated in the case of flat lying PFP/Cu(111) compared to flat lying Pn/Cu(111) and so as thin films these molecules are less structurally similar.  We do not see the same variety of phases as for Pn/Cu(111) \cite{smerdon2011monolayer} although sub-monolayer coverage is beyond the scope of this paper.  Symmetry arguments indicate that, at least to bilayer coverage, the film forms as one phase with six possible domains, as a result of two-fold mirror symmetry of PFP rows combined with three-fold rotational symmetry of the substrate. 

 -----------
 
 
 


\begin{acknowledgement}
Use of the Center for Nanoscale Materials, an Office of Science user facility, was supported by the U.S.  Department of Energy, Office of Science, Office of Basic Energy Sciences, under Contract No.  DE-AC02-06CH11357.  Primary support for this work was provided by the Department of Energy Office of Basic Energy Sciences (SISGR Grant No.  DE-FG02-09ER16109).  This work was also supported by Royal Society Research Grant RG130038.
\end{acknowledgement}
\section{SUPPORTING INFORMATION DESCRIPTION}
Supporting information is available free of charge at pubs.acs.org.
The information consists of additional scanning tunneling microscopy data.  The topics addressed are the identification of substrate lattice axes, the tilt of the molecules and the interaction with the tip.

\bibliography{sample}

\end{document}


\begin{abstract}
Here we provide additional data supporting the work detailed in the manuscript \emph{Monolayer and bilayer perfluoropentacene on Cu(111)}. Figures and discussion are presented regarding our identification of the Cu(111) sample axes,  and to illustrate the stable perfluoropentacene (PFP) ML formed at LT, demonstrating a semi-stable tilt about the long axis of the molecule and an apparent tip-induced tilt about the short axis of the molecule.
\end{abstract}

\maketitle
\tableofcontents
\section{Identification of Cu(111) primary axes}

\beginsupplement
\begin{figure}[h]
  \includegraphics{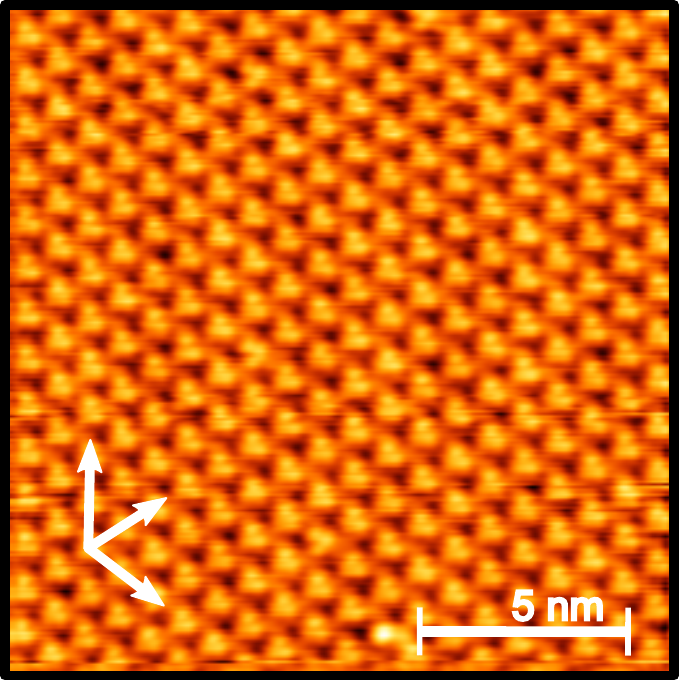}
  \caption{ A 15 $\times$ 15 nm scan of annealed C$_{60}$/Cu(111). The C$_{60}$ has adopted a p(4$\times$4) structure. The white vectors denote the C$_{60}$ and Cu(111) lattice axes. ($I_T=400\textrm{pA}, V_B=1.5$V).}
  \label{c60angle}
\end{figure}
As stated in the manuscript and as encountered in other works\cite{lagoute2004manipulation}, it was not possible to resolve individual atoms at the Cu(111) surface. Adsorption of C$_{60}$ atop Cu(111) has been well characterised and by annealing to 525 K, the p(4$\times$4) surface reconstruction\cite{pai2004structural} is formed.  We show this in Figure \ref{c60angle} in a measurement taken contemporaneously with those in the manuscript. The Cu(111) lattice axes can be determined by proxy as the C$_{60}$ NN directions lie parallel to those of the substrate. 

\section{Perfluoropentacene monolayer}
\begin{figure}[]
  \includegraphics{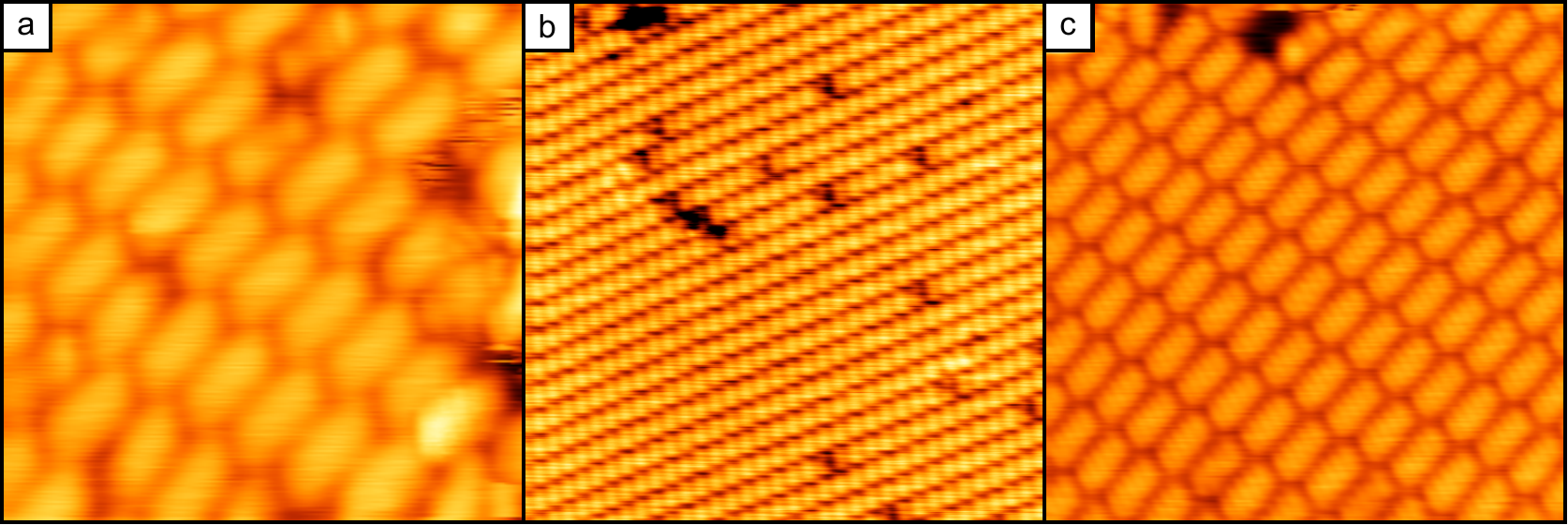}
  \caption{ a) A high resolution topograph of the monolayer showing the appearance of individual molecules as featureless rods ($I_T=100\textrm{pA}, V_B=1.5$V) b) large ML domains ($I_T=200\textrm{pA}, V_B=-3.0$V) c) At high negative biases we see the ML adopts a lobed appearance similar to that of the BL ($I_T=200\textrm{pA}, V_B=-3.0$V).}
  \label{monolayer}
\end{figure}

Stable ML formation is observed to occur at 50 K with large domains (Figure \ref{monolayer}(b)). Compared to the bilayer (BL) molecules we see a flat appearance at smaller biases (Figure \ref{monolayer}(a)). At defects within PFP ML some molecules appear to adsorb with a curved conformation or with bright end lobes (Figure \ref{monolayer}(c)). At higher negative biases we see the featureless appearance of ML PFP adopts a lobed appearance, similar to the BL.  This does not happen with Pn/Cu(111) at any bias reported \cite{smerdon2011monolayer}.

\section{Molecular tilt in the second layer}

A tip change occurs \nicefrac{1}{3} of the way into the down-scan shown in Figure 1(a), resulting in the BL molecules adopting two different appearances.  One is of a flat lying molecule that fits the HOMO model in Figure 2(a) and the other appearing as the tilted structure of Glowatzki \textit{et al.} at room temperature \cite{glowatzki2012impact}.  The tip change we observe coincides with the disappearance of a partially imaged PFP molecule from the BL; it seems reasonable to consider that this may have been adsorbed on the tip.

The effect of a molecule adsorbed on the tip is difficult to assess due to the unknown geometry of the arrangement \cite{chiutu2012precise,lakin2013recovering}.  However, in this instance, it produces an image very similar to that reported for PFP/Cu(111) by Glowatzki \emph{et al.} \cite{glowatzki2012impact} in their analysis of the `tilt' of the molecules.

\begin{figure}[]
  \includegraphics[width=16cm]{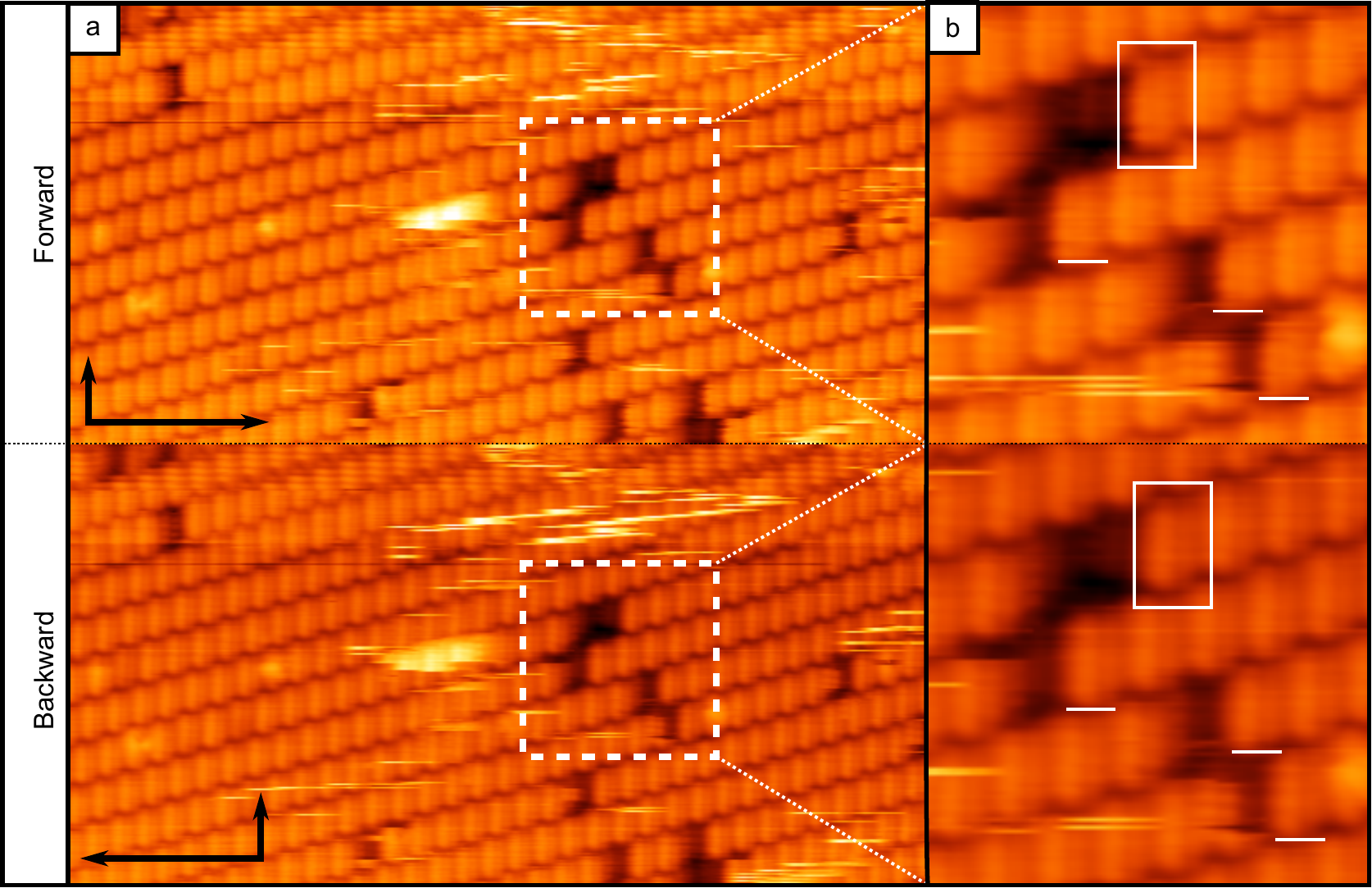}
  \caption{The fast and slow scan directions for each image are marked with a long and short arrow respectively; a) consistent BL tilt imaged in both forward and back scans; b) close up of the defected regions in (a); molecules with a defect adjacent to their normally raised edge appear flat flying in the forward scan but tilted in the backward scan ($I_T=100\textrm{pA}, V_B=-2.5$V)}
  \label{consistenttilt}
\end{figure}

\begin{figure}[]
  \includegraphics[width=16cm]{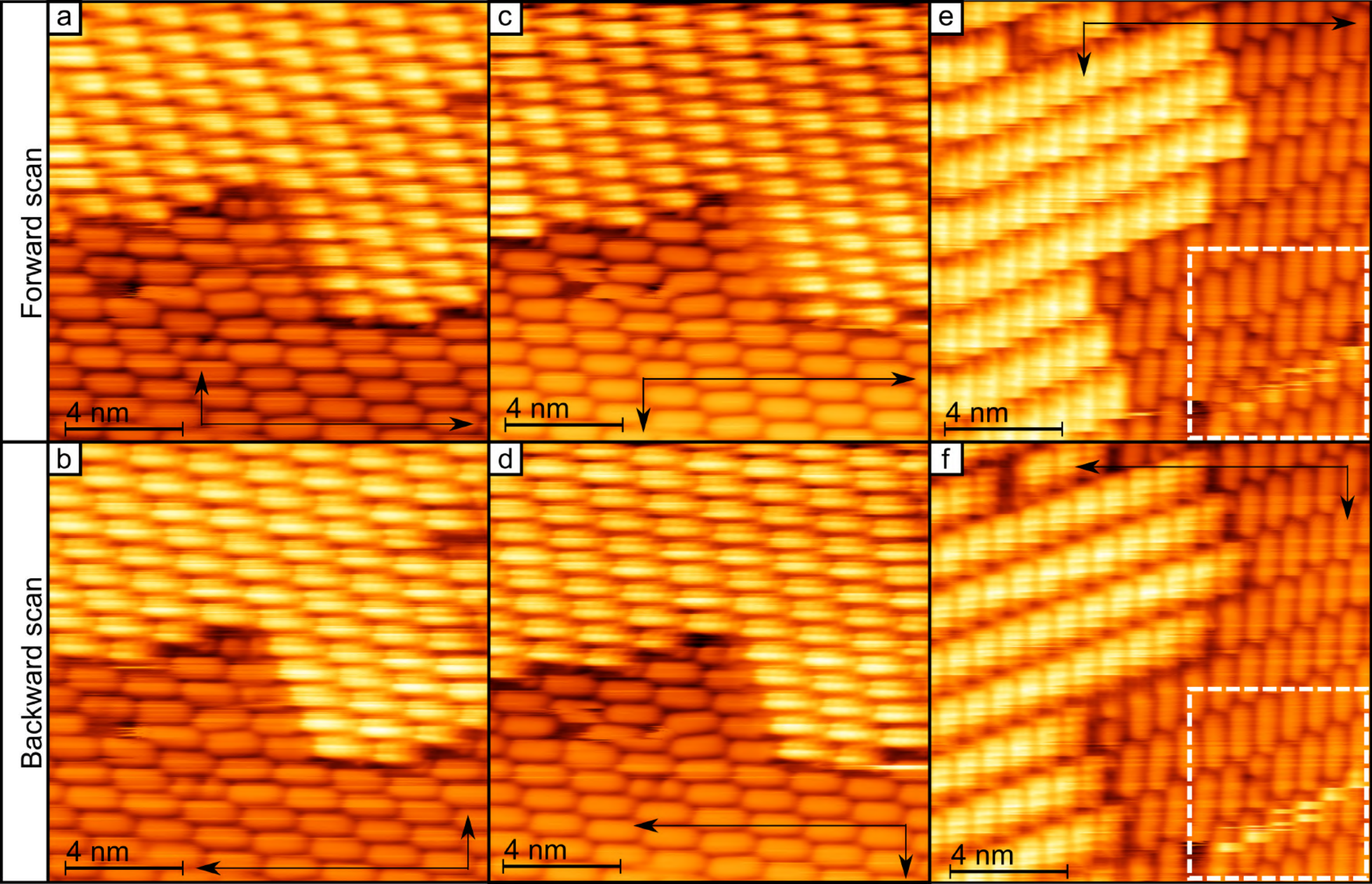}
  \caption{The fast and slow scan directions for each image are marked with a long and short arrow respectively; a) forward up; b)backward up; c) forward down; d) backward down; e) forward down, 90$^\circ$ rotated; f) backward down scan 90$^\circ$ rotated.}
  \label{scansbackforth}
\end{figure}

When attempting to verify the tilting structure reported by Glowatzki\textit{et al.} in the BL\cite{glowatzki2012impact}, we see instances of apparent tilting about the short and long axes of molecules.  The orientation and presence of this tilt changes dependent on the scanning motion of the tip and so seems likely to be driven by the tip-molecule interaction or a scanning artefact. Examples of this are presented in Figure \ref{consistenttilt} and Figure \ref{scansbackforth} for molecules seemingly tilting about their long and short axes respectively.

Figure \ref{consistenttilt}(a) shows a BL region with a tilt along the left edge of molecules within the BL in both forward and backward scans. The slow scan direction in both extends up the frame. This is consistent with the tilt reported by Glowatzki \textit{et al.} but is the only example of such in our entire dataset. For the majority of BL regions observed either the tip travel direction is seen to invert the apparent tilt of molecules, indicating either a tip-molecule interaction or an asymmetrical tip. In Figure \ref{consistenttilt}(b) we see within the same image that the appearance of BL molecules that are adjacent to defects changes based on the scan direction. 

Comparing forward and backward scans show in Figures \ref{scansbackforth}(a), \ref{scansbackforth}(b) and in Figures \ref{scansbackforth}(c) and \ref{scansbackforth}(d) we see different tilting about the long axis of the molecules dependent on the slow scan direction.

Inverting the fast and slow scan directions as in Figure \ref{scansbackforth}(e) and \ref{scansbackforth}(f) results in molecules seeming to tilt about their short axis. The tip-dependent tilting may stem in part from the staggering of the BL molecules. In Figure \ref{scansbackforth}(e) the tip travels from left to right; as it leaves one row it enters a depression between rows which has the effect of making the first line across a new molecule in the next row appear lower as the tip lags behind the surface contour.

\bibliography{sample}